# Height-Dependent Rotor Noise and Thrust in Urban Air Mobility: An Experimental Study

Qiyu Huang

Supervisor: Professor Mahdi Azarpeyvand

Cooperator: Jingwen Deng; Zishuo Lin

**ABSTRACT**

*The present study investigates the aerodynamic and aeroacoustic characteristics of a propeller operating under varying rotational speeds (RPM) and heights (H), with a particular focus on the effects of upstream obstruction modelled as a tall building. Unlike previous studies that primarily examined rotor noise under axial inflow conditions, this work explores how vortex shedding and flow ingestion from different elevations influence rotor performance and noise emissions. Experiments were carried out in an anechoic wind tunnel, where a tall cylinder was positioned above the propeller to replicate real-world obstruction scenarios. Results revealed that lower propeller heights led to increased broadband noise due to intensified turbulence interactions and reduced aerodynamic efficiency, while higher positions improved thrust performance and mitigated noise effects under certain conditions. The findings contribute to understanding noise sources in eVTOL propulsion systems and provide insights for optimizing propeller placement to enhance aerodynamic efficiency and noise reduction in urban environments.*

## 1 Introduction

In recent years, significant attention has been directed toward electric Vertical Take-Off and Landing (eVTOL) vehicles as a promising solution for Urban Air Mobility (UAM), primarily due to their potential to reduce congestion and enhance urban transportation efficiency. However, noise generated from eVTOL rotors remains a major challenge, impacting public acceptance and operational viability within densely populated areas. Rotor noise is typically divided into tonal and broadband components, each associated with different physical mechanisms. Tonal noise primarily originates from deterministic blade loading variations, rotational harmonics, and blade thickness effects, contrasting with broadband noise which is attributed to stochastic mechanisms including turbulence ingestion, unsteady aerodynamic interactions, and turbulent boundary layer flows around the blades.

Despite substantial progress, the interplay between rotor operational parameters—rotor height (*H*) relative to urban structures, rotational speeds (RPM), and the resultant acoustic and aerodynamic impacts remains inadequately explored. This study aims to address this research gap through a systematic experimental investigation of aeroacoustic and aerodynamic measurements to explore the interactions resulting from varying propeller heights and rotational speeds in the presence of upstream cylindrical obstructions, specifically focusing on rotor aerodynamic performances (quantified by thrust coefficient, $C_T$) and aeroacoustic characteristics (e.g., broadband modulation, tonal harmonics, and overall sound pressure level). This scenario effectively simulates the real-world urban operational conditions encountered by eVTOL vehicles.

Recent experimental research has particularly emphasized the complexity of these noise sources when rotors operate under non-axial inflow conditions, typically during transition phases of flight, hover, and when interacting with urban infrastructures such as buildings and landing platforms. Hanson et al. demonstrated that negative tilting angles during descent and hover conditions significantly enhance the broadband noise component due to increased re-ingestion of rotor self-generated turbulence through extensive experiments. Similarly, Celik et al. highlighted the substantial impact of rotor arrangement, specifically in tandem configurations, on both tonal and broadband acoustic emissions, pointing to the necessity of careful aerodynamic configuration in reducing overall noise signatures.

Moreover, vortex shedding—another critical aeroacoustic mechanism—significantly contributes to tonal noise emission from bluff bodies and propellers placed in proximity. Research conducted by Chen et al. on vibrissa-inspired cylinder shapes illustrated how bio-inspired geometries can alter vortex shedding behaviors and remarkably attenuate tonal noise by reducing coherent vortex strength. This indicated that controlling the coherence and strength of vortex shedding near rotor structures could effectively mitigate noise—an insight highly relevant for eVTOL designs.

The phenomenon of boundary layer ingestion and turbulence ingestion noise also plays a pivotal role in broadband noise generation in urban air mobility vehicles. Recent experimental work by Guérin et al. revealed that boundary-layer ingestion notably elevated broadband noise levels, highlighting that low Reynolds-number and low-Mach number propeller operations, typical of urban air taxis, were especially susceptible to these effects. Such insights are critical for informing aerodynamic designs that mitigate these noise generation mechanisms.

Additionally, research carried out by Gan et al. on multi-rotor aircraft emphasized the importance of broadband noise modulation and its time-varying characteristics, demonstrating significant fluctuations in noise levels due to azimuthal phasing and rotor interactions. Their findings suggest that noise modulation significantly impacts human perception, revealing the necessity of incorporating time-resolved acoustic analyses and modulation-sensitive noise mitigation strategies such as synchrophasing into eVTOL design and operation.

Further complexities arise from interactions between rotors and adjacent structural surfaces, including the ground and buildings. Kan et al.  emphasized the significant aerodynamic and acoustic influences of ground effect on rotorcraft, revealing that proximity effects alter both performance characteristics and acoustic signatures. Similarly, experimental findings by Qin et al. and Yangzhou et al.  on turbulence-ingesting propellers underlined that interactions with turbulent wakes drastically influenced noise emissions and aerodynamic efficiency, further complicating acoustic mitigation efforts in urban environments.

By integrating synchronous aeroacoustic and aerodynamic measurement techniques, this research seeks to provide comprehensive analysis of how turbulence ingestion, wake alignment, and structural proximity modulate rotor noise emissions and aerodynamic efficiency. These findings aim to deliver valuable insights into noise reduction strategies, aerodynamic optimization, and practical considerations for future eVTOL designs operating within complex urban aerodynamic environments.

# 2 Methodology

## 2.1 Experimental Methodology

Experimental investigations into the aeroacoustic characteristics of a two-bladed rotor were conducted at the Aeroacoustics Facility of the University of Bristol, within an anechoic chamber measuring 7.9 m × 5.0 m × 4.6 m (L × W × H), with a cut-off frequency of 160 Hz. A commercially available two-bladed APC propeller with a diameter of 0.3048 m (12 inch) was mounted 0.6 m downstream from the nozzle exit of a closed-circuit, open-jet wind tunnel. This configuration enabled steady, low-turbulence inflow conditions essential for reliable aeroacoustic measurements.

Acoustic data were recorded using 62 GRAS 40PL microphones, strategically distributed across two arrays to capture the directional characteristics of the radiated sound. The top array was positioned 1.5 m above the propeller hub, spanning polar angles from 40° to 145°, while the side array was located 2 m from the hub, covering azimuthal angles between 301° and 340°. The microphone at 340° was disconnected during testing due to signal integrity issues. Figure 1 illustrates the schematic layout of the experimental apparatus, highlighting the axial positioning of the propeller within the chamber to ensure minimal boundary interference and uniform flow exposure. Figure 2 below provides a detailed view of the microphone arrangement, showing how both vertical and horizontal dispersion patterns were captured through this comprehensive setup.

Such spatial coverage is critical for accurate resolution of directivity patterns and broadband noise content.

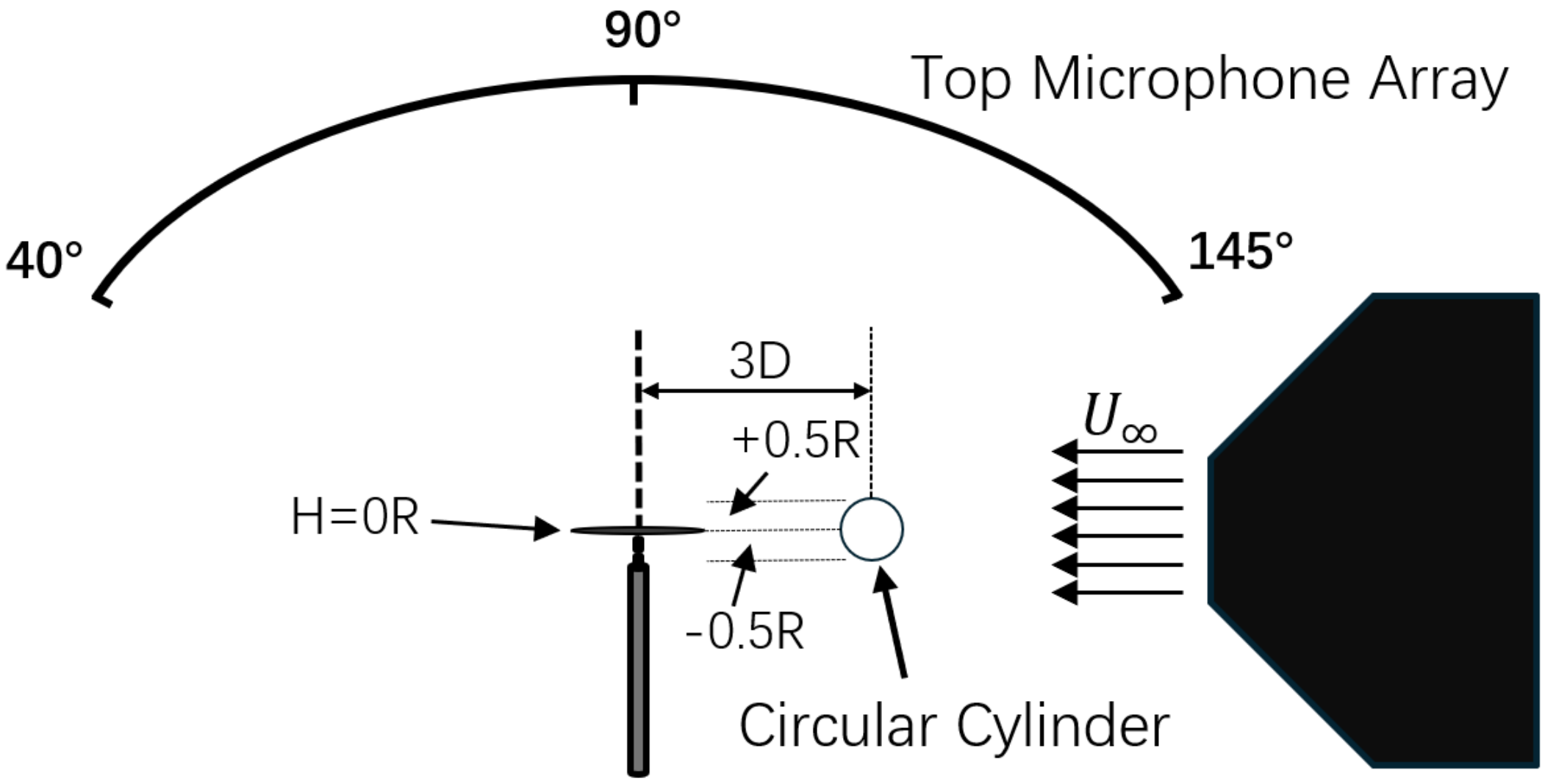

Figure 1: Schematic layout of the experimental apparatus, showing the axial positioning of the propeller within the anechoic chamber

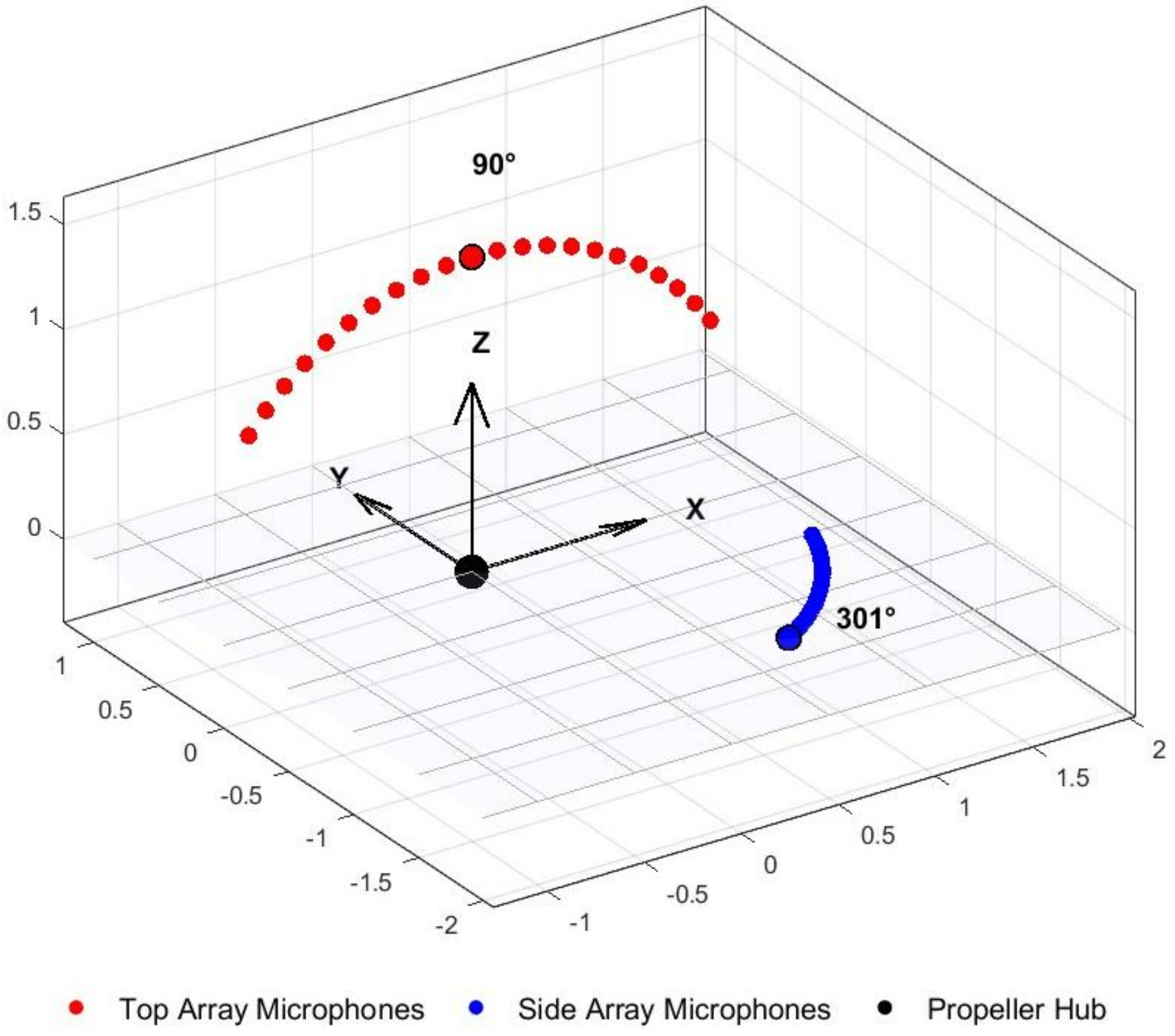

Figure 2: Microphone Array Arrangement. Top array microphones which were shown in red points varied from 40˚ to 145˚ (from left to right, placed every 5˚). Side array microphones shown in blue points varied from 301˚ and 340˚ (from left to right placed every 1˚)

## 2.2 Aeroacoustic Analysis

Sound Pressure Level (SPL) was computed using Welch's method, implemented by segmenting the acoustic signals with a 50% overlap and employing a Hamming window function, leading to a frequency resolution of 1 Hz, given the sampling frequency (65536 Hz) and window size utilized. SPL was calculated at each frequency by:

$$SPL=10\log_{10}\left(\frac{\phi_{pp}}{p_{ref}^2}\right) \tag{1}$$

where $\phi_{pp}$ is the power spectral density of pressure fluctuations, and $p_{ref}$=20 μPa represents the reference sound pressure.

Overall Sound Pressure Level (OASPL) was computed by integrating the acoustic energy spectrum across the frequency range of interest:

$$OASPL=10\log_{10}\left(\frac{\int PSD(f)df}{p_{ref}^2}\right) \tag{2}$$

## 2.3 Aerodynamic Analysis

Aerodynamic performance was evaluated using thrust measurements obtained through a calibrated load cell. These measurements were normalized to obtain non-dimensional thrust coefficients $C_T$ using the expression:

$$C_T=\frac{T}{\rho(\Omega R)^2 A} \tag{3}$$

where $T$ denotes the measured thrust, $\rho$ the air density, Ω the rotor angular velocity, $R$ the rotor radius, and $A$ the rotor disc area. The aerodynamic analysis was systematically conducted at varying propeller heights and RPM settings, in conjunction with the corresponding acoustic analysis, to comprehensively examine their influences on rotor performance.

# 3 Results & Discussion

## 3.1 Aerodynamic Performance

By normalizing thrust, $C_T$ isolates aerodynamic performance independent of specific thrust level or rotor size. The inflow velocity was varied from 8 ms$^{-1}$ to 24 ms$^{-1}$. The experimental methodology ensured consistent data collection at each speed and height, allowing direct comparison of $C_T$ across the configurations. All other parameters (RPM, blade pitch, etc.) were held constant, so any differences in $C_T$ are attributable to aerodynamic effects of varying incoming velocities and propeller heights. Clearly shown in Figure 3, at low incoming velocities, there was little difference in thrust coefficient between $H$=0.5$R$ and $H$=0$R$. This indicated that when the inflow speed is small, mounting the propeller above the centerline does not yet confer a notable advantage over being mounted in-plane. $H$=-0.5$R$ case at low speed showed only a slight improvement relative to the other two. In practice, $C_T$ does also vary with rotor speed due to Reynolds number and compressibility effects. Deters et al. observed that for small-scale propellers, increasing Reynolds number (via higher RPM) yielded improved performance: $C_T$ increased and power coefficient decreased as Re increased. In contrast, at lower RPM, blade Reynolds numbers fall, and airfoil efficiency drops. A slow-turning rotor produces less thrust per rotation due to low-Reynolds penalties. Additionally, at very high RPM, compressibility can play more significant role which might incur minor losses from shock formation and drag rise and lead to reducing $C_T$.

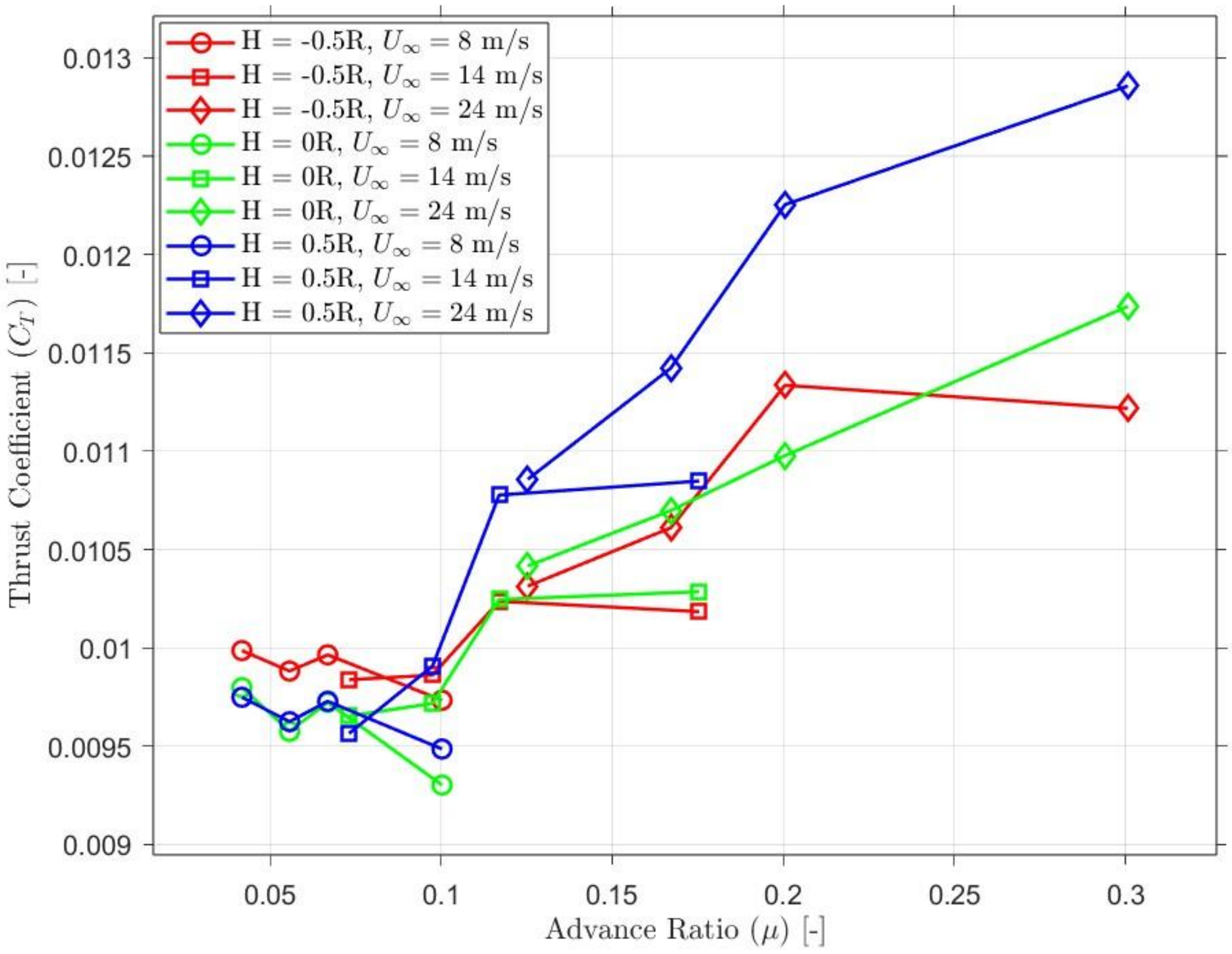

Figure 3: Thrust Coefficient at different advance ratio. Blue lines represent $H = 0.5R$, red lines represent $H = -0.5R$ and green lines represent $H = 0R$.

Denoted by Figure 3 above, at low incoming velocity (8 $ms^{-1}$), $H$=-0.5$R$ showed that a slightly higher $C_T$ could be explained by which pointed out that a significant thrust augmentation will occur as the ground gets closer. This increase is due to the well-documented ground effect phenomenon. The ground reduces the induced velocities below the rotor, reducing induced drag and thereby increasing thrust efficiency. The rotor effectively utilizes a cushion of high-pressure air trapped beneath it, leading to greater thrust output at a given advance ratio. For 8 $ms^{-1}$ incoming flow, blue lines which represent $H$=0.5$R$ showed that as the propeller being further from the ground, thrust coefficients are consistently lowest across nearly all advance ratios tested. This result aligns with literature, which indicates that aerodynamic efficiency of a rotor generally decreases as the rotor is moved away from the ground plane.

The observed increasing trend in $C_T$ with $\mu$ is consistent with classical rotorcraft aerodynamics in edgewise flow. In classical rotor aerodynamics, increasing $\mu$ leads to a reduction in induced velocity, allowing more efficient inflow into the rotor disk. This effect, often referred to as translational lift, occurs as the rotor moves from operating in its own wake at low $\mu$ to interacting with cleaner freestream air. Wake convection also plays a crucial role in modifying rotor inflow characteristics. In low $\mu$ conditions, tip vortices remain near the rotor, increasing blade-vortex interaction losses and disturbing inflow uniformity. As $\mu$ increases, the rotor wakes tilts rearward and convects downstream, reducing vortex recirculation and allowing more stable and uniform inflow into the rotor disk. This wake stabilization effects enhances rotor efficiency. Johnson also demonstrated that blade aerodynamics contribute to the variation of $C_T$. At low advance ratio, the rotor operates in an inflow condition where some blade sections, particularly near the root, may experience higher angles of attack and separated flow, leading to inefficient lift generation. As $\mu$ increases, the local inflow angles adjust, favoring improved aerodynamic performance. The advancing blade encounters a higher relative velocity and benefits from enhanced aerodynamic efficiency, while the retreating blade experiences reduced adverse effects due to wake convection. The influence of RPM on these trends is also vital. At higher RPM, the rotor operates at a higher thrust efficiency, so the relative increase in $C_T$ with $\mu$ is smaller compared to lower RPM cases. At lower RPM, the rotor operates at a less favorable Reynolds number, and forward velocity compensates for these inefficiencies, resulting in a more pronounced increase in $C_T$ with $\mu$.

In conclusion, the observed increasing trend in $C_T$ with $\mu$ is consistent with the aerodynamics of the classical rotorcraft in the flow in the edge direction. Translational lift, wake convection, and improved blade aerodynamics collectively contribute to this behavior. These findings reinforce the importance of considering wake dynamics and inflow characteristics when evaluating rotor performance at low to moderate advance ratios. Future research shall carry out detailed flow analysis around the rotor disc so that flow visualization can further check the wake interaction differences for different propeller settings.

## 3.2 Aeroacoustic Performance:

In the experimental setup, a horizontal cylinder positioned 3D upstream of the propeller where D is the diameter of the cylinder introduces a Kármán vortex street, characterized by periodic vortex shedding at frequencies determined by the Strouhal number, cylinder diameter and flow velocity. These shed vortices and the turbulent wake from the cylinder interact with the downstream propeller blades, leading to complex aeroacoustics phenomena involving both Blade-Wake Interaction (BWI) and Blade-Vortex Interaction (BVI) noise mechanisms. In Figure 4 below, the small peaks around the blade passing frequency (BPF) in low frequencies were referred to "Haystacking" which were detailly described in. In their investigation of boundary layer ingestion (BLI) configurations, haystacking was defined as the organization of broadband turbulence ingestion noise into humps near the BPF and was found to intensify with increased propeller thrust and reduced tip gap, both of which increase blade-to-blade correlation with large-scale turbulent eddies.

Figure 4 presents the SPL spectra for the three vertical propeller positions. While each case exhibits both tonal components (BPF harmonics) and broadband noise, a consistent pattern emerges across all heights: the 8 $ms^{-1}$ incoming flow case produces more prominent broadband humps near the BPF, a signature trait of the haystacking effect. The vortex shedding frequencies identified in the SPL spectra, characterized by small side peaks around the blade passing frequency, exhibit close agreement with theoretical predictions based on the Strouhal number. While minor deviations were observed between the experimental vortex shedding frequency and the canonical estimate from $f_s=(St\cdot U_\infty)/D$, where St represents the Strouhal number which typically was set into 0.2, $U_\infty$ represents the incoming velocities and $D$ represents the characteristic length (in this case, $D$ represents the diameter of the cylinder). These differences are within acceptable boundaries. This is expected, as the Strouhal number is known to vary slightly with Reynolds number, turbulence level, and cylinder aspect ratio, and is not strictly constant at 0.2 under all conditions. In practical wind tunnel environments, especially those involving complex inflow or blockage effects, the local Strouhal number may deviate. Thus, the experimental evidence remains consistent with the expected vortex shedding behavior, validating the association between these spectral features and blade-vortex interactions.

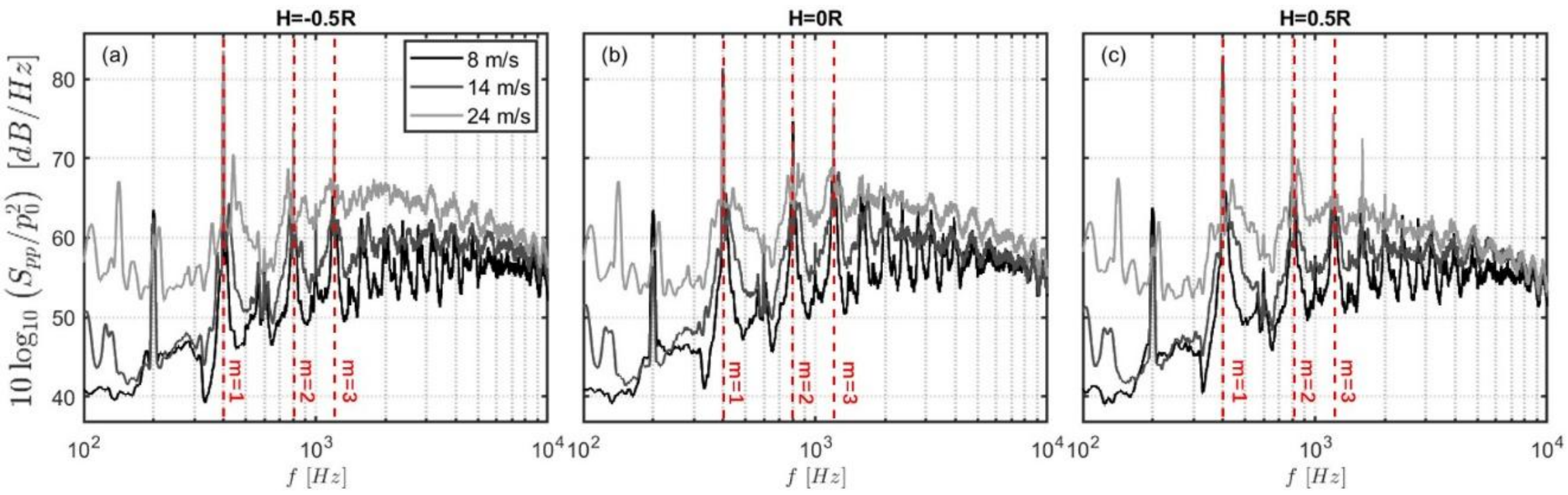

Figure 4: SPL graph for 12000RPM. (a) represents for $H = -0.5R$, (b) represents for $H = 0R$, (c) represents for $H = 0.5R$. Red lines for BPF and its harmonics from the first to the third harmonic.

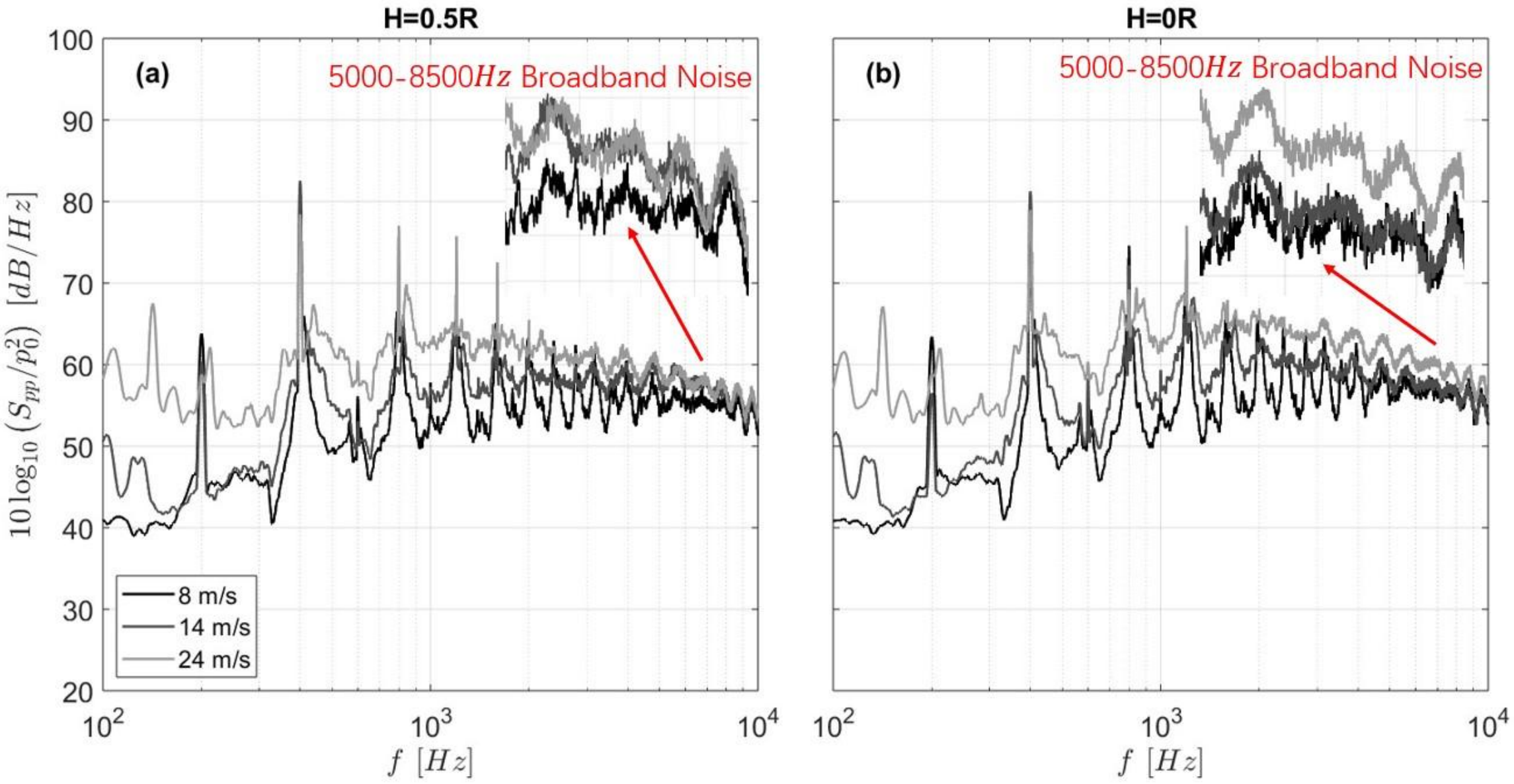

Figure 5: Zoomed-in SPL spectra for (a) $H = 0.5R$, (b) $H = 0R$ highlighting broadbandnoise behavior between 5 kHz to 8.5 kHz. (a) Overlap between 14 ms$^{-1}$ and 24 ms$^{-1}$ indicates modulation saturation at high velocities. (b) Overlap between 8 ms$^{-1}$ and14 ms$^{-1}$ suggests coherence-driven modulation effects.

Haystacking patterns are especially notable when comparing 8 ms$^{-1}$ to 14 and 24 ms$^{-1}$. While the higher velocity cases retain strong tonal content, their spectra beyond the third harmonic are smoother, lacking the modulated hump structures. This behavior can be understood within the framework of edgewise blade-vortex interaction and the haystacking mechanism proposed by Zaman et al. The authors showed that in turbulence ingestion scenarios, repeated blade encounters with coherent eddies lead to broadband noise reorganization into humps near BPF harmonics.

Comparing the SPL spectral of 8 $ms^{-1}$ cases for all propeller heights in this research to 14 and 24 $ms^{-1}$ cases, it can be easily found that the haystacking effects were more obvious after the third harmonics in 8 $ms^{-1}$. To better illustrate the detailed differences in broadband noise modulation and tonal harmonics between the $H$=0.5$R$ and $H$=-0.5$R$ configurations, zoomed-in spectra focusing on critical frequencies around the blade passing frequencies (BPF) are presented in figure 5. In low incoming velocity conditions, where the turbulence time scale becomes long relative to the blade-passing period, increasing the probability that multiple blades interact with the same turbulent structure. This will lead to a strong blade-to-blade correlation in turbulence ingestion, which has been identified as a key mechanism responsible for haystacking features in computational and experimental studies. In contrast, at higher inflow velocities (14 and 24 $ms^{-1}$), the modulation diminishes, and the SPL spectra beyond the third harmonic become smoother and more continuous. This may because the turbulent eddies convect more rapidly through the rotor disk, reducing coherence and making each blade's interaction independent.

Additionally, the presence of a horizontal cylinder upstream, shedding a Kármán vortex street, introduces further periodicity into the inflow for the propeller. These vortex structures, particularly coherent at lower speeds, serve as a forcing mechanism that enhances blade-vortex interaction (BVI) and blade-wake interaction (BWI) noise components. Yangzhou et al. demonstrated that the ingestion of such wake structures by a downstream propeller significantly alters the aeroacoustic field, with more distinct spectral features appearing when the vortical content is strong and coherent.

To ensure a more focused evaluation of the height-dependent broadband noise behavior, the Overall Sound Pressure Level (OASPL) analysis was confined to a filtered frequency range of 5 kHz–8.5 kHz. This selection was guided by spectral observations from the zoomed-in SPL plots of $H$=0.5 $R$ and $H$=0 $R$ configurations (Figure 5), which clearly indicate that acoustic curve convergence and modulated broadband amplifications occur most prominently within this frequency band. Specifically, the 8 $ms^{-1}$ and 14 $ms^{-1}$ spectra at $H$=0 $R$ display overlapping patterns in the broadband region beyond the third Blade Passing Frequency (m=3 red lines in figure 4) harmonics, especially between 5 kHz and 8.5 kHz, while a similar behavior is observed for the 14 $ms^{-1}$ and 24 $ms^{-1}$ cases at $H$=0.5 $R$. This focused approach allows for a more accurate interpretation of broadband noise generation and acoustic sensitivity modulation across different inflow velocities and propeller positions.

An intriguing observation raised when comparing OASPL (figure 6) trends at H=0R and H=0.5R; at $H$=0$R$, the 8 $ms^{-1}$ and 14 $ms^{-1}$ cases exhibit nearly overlapping OASPL levels across top-array directions, while at $H$=0.5$R$, the 14 $ms^{-1}$ curve closely follows that of 24 $ms^{-1}$. These anomalies deviate from the expected monotonic increase of OASPL with inflow velocity where Hanson et al. showed that increased turbulence intensity and integral length scale led to stronger interactions with the propeller, resulting in higher OASPL in the acoustic far-field. This observation suggests a height-dependent modulation in how turbulent structures interact with the rotor disk.

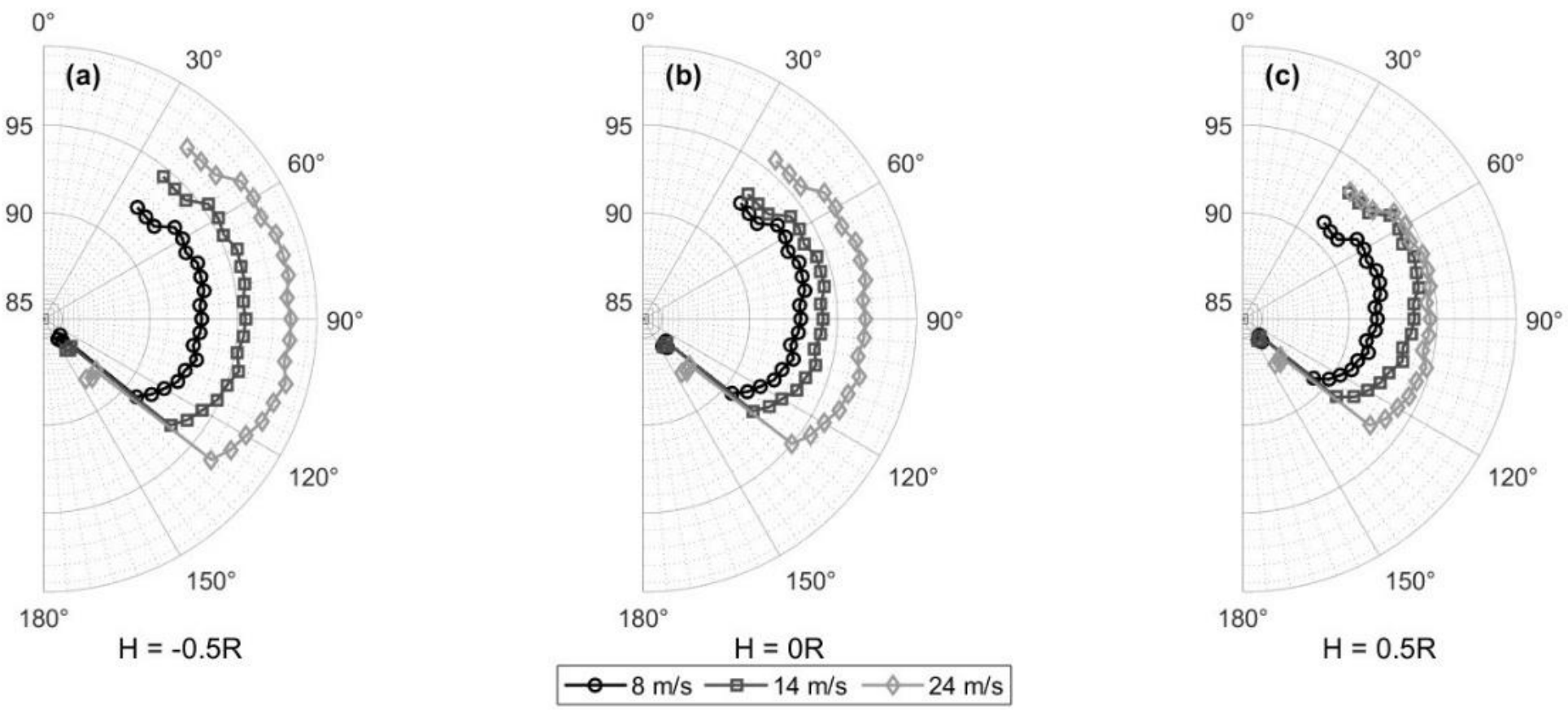

Figure 6: Overall Sound Pressure Level for all $H$ cases for different incoming velocities. Capturing for 5 kHz to 8.5 kHz which broadband noise is the dominant noise source.

At $H$=0R, the rotor lies directly in the central axis of the cylinder's wake, where vortices are most coherent and symmetrically aligned with the vortex structure. This configuration enables strong blade-to-blade loading correlation in the 8 $ms^{-1}$ case, enhancing haystacking and producing broadband noise levels that are comparable to the more energetic but less coherent 14 $ms^{-1}$ case. This reflects a balance between coherence-driven modulation and energy-driven amplification, which leads to converging OASPL magnitudes. At $H$=0.5$R$, the propeller is displaced vertically above the vortex core trajectory. Here, the 14 $ms^{-1}$ case interacts with partially decohered yet still energetic vortex structures, creating a strong acoustic response dominated by broadband noise. Since the 24 $ms^{-1}$ case adds further energy without introducing significant changes to turbulence structure, the OASPL increase is modest, and the two curves appear nearly superimposed. This suggests that the acoustic sensitivity saturates at this height once a threshold level of turbulence intensity is reached, especially when flow coherence is reduced. In contrast, the behavior at $H$=-0.5$R$ is markedly more regular. The OASPL differences between 8, 14, and 24 ms$^{-1}$ are clearly distinguished and increase monotonically with velocity across all polar angles. This can be attributed to the rotor being positioned below the wake core, where it ingests less organized, more uniformly randomized turbulence. The absence of strong vortex-rotor alignment at this height minimizes haystacking effects and blade-to-blade coherence, causing the OASPL to scale primarily with turbulence energy. As a result, the acoustic response becomes more linear and predictable, in contrast to the transitional behavior observed at $H$=0$R$ and $H$=0.5$R$.

This interpretation is consistent with findings from recent wake-ingestion studies, which emphasized that acoustic non-linearity arose when coherent structures were partially ingested by the rotor, especially in edgewise or non-axial flow. The current data further support that height-dependent vortex alignment plays a critical role in dictating not just spectral content but also total radiated energy.

Furthermore, Figure 4 reinforces the interpretation that $H$=-0.5$R$ exhibits reduced haystacking behavior compared to $H$=0$R$ and $H$=0.5$R$. In the -0.5$R$ configuration, particularly at 8 ms$^{-1}$, the broadband hump around the first BPF harmonic after the third harmonic remains subdued and does not significantly exceed the 14 ms$^{-1}$ case. This spectral behavior is consistent with the earlier observation that the propeller at -0.5$R$ is positioned outside the core region of the upstream vortex street, thereby ingesting more randomized turbulence with less coherence and fewer blade-to-blade correlations. As a result, the noise response becomes less modulated and more energy-driven, aligning with the predictable monotonic increase in OASPL seen at this height.

In contrast, at $H$=0$R$ and 0.5$R$, the 8 ms$^{-1}$ SPL curves show stronger modulation and broadband amplification adjacent to the harmonics, which aligns with the hypothesis of enhanced interaction with coherent structures from the cylinder’s wake. The increased presence of quasi-periodic spectral content at these heights underscores the influence of wake alignment and inflow structure on the noise generation mechanism, particularly in edgewise flow conditions. These findings collectively validate the role of rotor height in dictating not only the spatial distribution of acoustic energy but also the spectral modulation patterns associated with turbulence ingestion and blade-wake interactions.

Figure 7 showed OASPL for side-array microphones which measured the far-field noise from 301° to 339°, providing detailed insight into how propeller height modulates angular noise patterns under edgewise flow. All three height cases showed a broad OASPL dip near 320°, corresponding to the downstream axis of the propeller wake, but the depth, shape, and inter-velocity spacing across polar angles reveal important differences. At *H*=-0.5*R*, the OASPL curves show a consistently monotonic spacing between 8 $ms^{-1}$, 14 $ms^{-1}$, and 24 $ms^{-1}$ cases across all angles, indicating that the rotor disk is ingesting randomized turbulence with minimal coherence. This yields a more isotropic and energy-driven acoustic response, in agreement with the SPL spectra (Figure 4), where modulation was weak and haystacking was minimal.

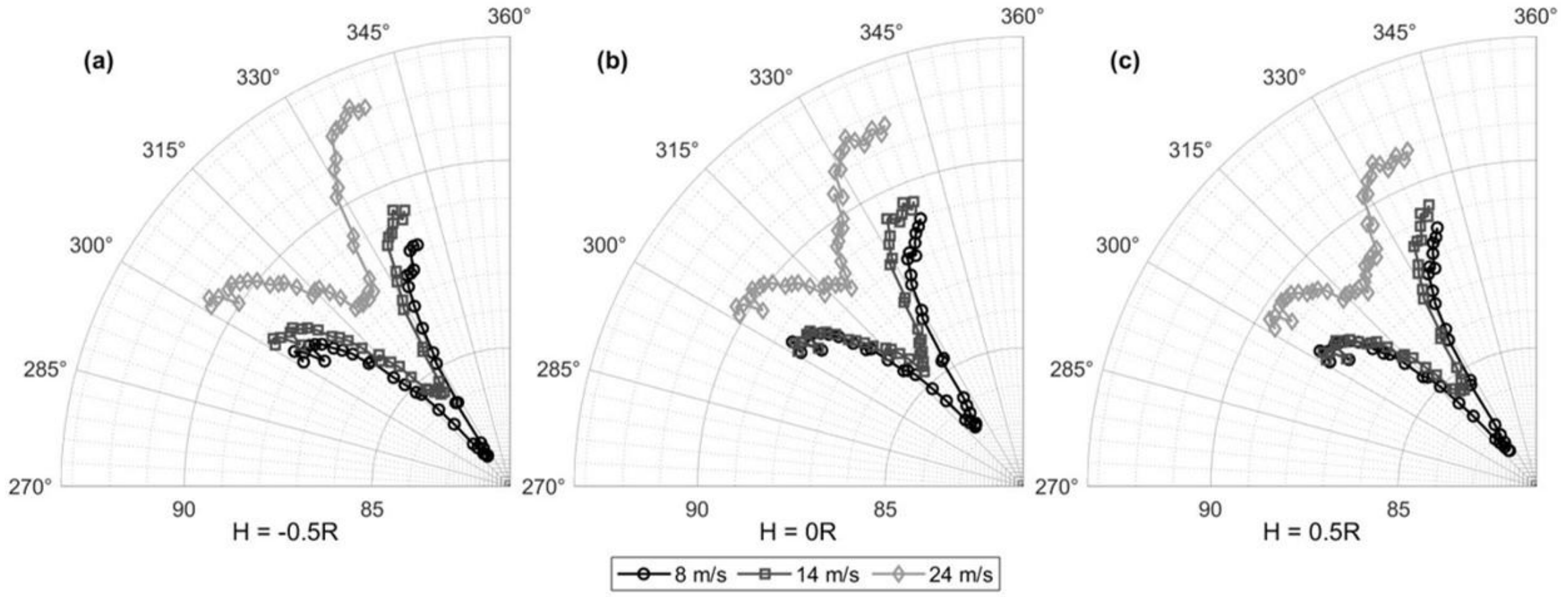

Figure 7: Side-Array Microphone OASPL for different propeller heights.

In contrast, both *H*=0*R* and *H*=0.5*R* show very similar side-array OASPL shapes and magnitudes, with nearly identical dip profiles and curve flatness beyond 320°. However, a closer examination of the early angular region (305°–312°) reveals that *H*=0*R* exhibits a larger pointwise difference between the 8 $ms^{-1}$ and 14 $ms^{-1}$ curves compared to *H*=0.5*R*. This suggests that while the total energy and radiation pattern are comparable, the degree of spectral modulation due to coherent vortex ingestion is greater at *H*=0*R*. Here, the rotor lies directly in the core of the Kármán vortex street and thus encounters more synchronized, quasi-periodic structures, particularly in the 8 $ms^{-1}$ case. These structures enhance blade loading fluctuations and increase side-radiated acoustic energy, resulting in greater OASPL divergence with increasing velocity.

At *H*=0.5*R*, the rotor is positioned above the core of the wake, where coherent eddies are still present but have likely diffused or tilted, producing weaker blade-to-blade coherence. This results in the 14 $ms^{-1}$curve remaining closer to 8 $ms^{-1}$ at the same angles, even though both heights exhibit similar total radiation.

Taken together, these findings demonstrate that small-scale variations in wake alignment and coherence have measurable impacts on local acoustic radiation, even when global OASPL shapes appear similar. The greater early-angle divergence at $H=0R$ confirms that the propeller's position within the wake's most coherent region plays a key role in haystacking and side-radiated noise modulation.

## 3.3 Validations:

### 3.3.1 Proper Orthogonal Decomposition (POD)

Time-resolved flow visualization should complement the acoustic measurements. Applying Particle Image Velocimetry (PIV) or stereoscopic PIV in the plane of the rotor and wake would allow direct observation of how vortices shed by the cylinder convect and interact with the rotor at different heights. Such data would validate the inferred mechanisms (e.g., confirming blade-to-blade coherence when the haystacking was observed). Based on temporal-only data which was the result of the testing in the anechoic wind tunnel in University of Bristol, Proper Orthogonal Decomposition (POD) was carried out for a first trial of extracting the most energetic structures in the fluid. POD is a data-driven technique designed specifically to extract coherent spatial-temporal structures from experimental or numerical datasets which is considered to be an alternative method for PIV. Its theoretical foundation requires a complete spatial-temporal data matrix, structured as spatial locations×time snapshots. Typical POD data matrix is defined as: $X(\mathbf{x},t)$, where $\mathbf{x}$ denotes spatial positions, and t denotes time snapshots. POD modes are obtained by solving the eigenvalue decomposition problem for the spatial correlation (covariance) matrix:

$$R = X X^T, \quad R_{ij} = \langle u(\boldsymbol{x}_i, t) u(x_j, t) \rangle_t$$

In this work, acoustic data were gathered from strategically positioned top and side microphone arrays, subsequently arranged into a snapshot matrix with spatial points represented by microphone positions and temporal points represented by time-series data. Our dataset consisted only of temporal information (i.e., without genuine spatial information), the fundamental assumptions behind POD break down. By mistakenly set the positions of the 62 GRAS 40PL microphones as spatial data, POD modes inherently reflect spatial structures evolving in time. This artificial spatial definition for the microphone positions as spatial points, constructed a snapshot matrix as follows:

$$X = \begin{pmatrix} p_1(t_1) & p_1(t_2) & \cdots \\ p_2(t_1) & p_2(t_2) & \cdots \\ \vdots & \vdots & \ddots \end{pmatrix}$$

Mathematically, this matrix is a valid numerical input for Singular Value Decomposition (SVD). SVD was applied to this mean-centered snapshot matrix, mathematically expressed as:

$$X_{centered} = U\,S V^T \tag{4}$$

SVD was used to extract spatial modes (**U**), temporal coefficients (**V**), and associated mode energies from the singular values (**S**). Despite the artificial definition of spatial dimensions—microphone positions rather than actual flow-field points—SVD effectively decomposed the dataset, facilitating the identification of spatial distribution patterns in the acoustic data. While acknowledging that the spatial POD modes derived from acoustic pressure measurements do not fully represent actual fluid flow structures, the approach provided insightful preliminary visualization of acoustic-related spatial coherence. In the core part of the POD analysis, a 4/3 coefficient distance correction (from (2 m) to (1.5 m)) was applied to the side array, and a 0.2 second segment was extracted from the middle of the entire time series for analysis. During the data preprocessing stage, the code used a frequency domain bandpass filtering method (implemented with adaptive masking) to extract pressure signals within the 5000–8500 Hz range. The core POD analysis constructed a spatial-temporal data matrix $X(\mathbf{x},t)$, performed time-averaging de-centering, and applied SVD. The results showed that the code calculated the energy contribution rate and cumulative energy distribution of each mode, visualized the spatial distribution of dominant modes through polar plots for the top array and linear plots for the side array, and drew time variation curves for the first three modes. In the modal reconstruction phase, the code selected a sufficient number of modes based on a 90% energy threshold, reconstructed the original signal, and calculated the Frobenius norm error. Finally, a time-frequency analysis of the modal temporal coefficients was performed using an optimized STFT method ($2^9$ window length, 50% overlap, $2^{11}$ point FFT), revealing the spectral characteristics of the main modes and their distribution in the frequency domain, providing key insights for understanding the spatial coherent structures and propagation characteristics of rotor noise.

### 3.3.2 Inverse Lighthill-Acoustic-Analogy for Flow field Simulation

This study employed an inverse flow-field reconstruction method based on acoustic pressure data obtained from a microphone array, combined with Lighthill's acoustic analogy. The motivation for this approach stemmed from experimental constraints that prevented direct spatial velocity field measurements—such as those provided by Particle Image Velocimetry (PIV)—which are essential for analyzing wake and vortex behavior with high temporal resolution. As an alternative, we explored the use of far-field acoustic data and applied Proper Orthogonal Decomposition (POD) to extract dominant spatio-temporal acoustic modes. The extracted modes were then used in conjunction with Lighthill's acoustic analogy to approximate flow features based on the spatial structure and gradients of the acoustic field. Similar inverse methodologies have been preliminarily explored in aeroacoustics to characterize source regions. Inspired by these ideas, we attempted to extend this framework to infer flow structure coherence in edgewise inflow rotor conditions. According to Lighthill's acoustic analogy, the physical mechanisms of aerodynamic noise generation explicitly depend on the nonlinear velocity terms ($\rho u_i u_j$) and turbulent viscous dissipation ($\tau_{ij}$), rather than solely on far-field acoustic pressure fluctuations ($p'$). Mathematically, Lighthill's equation is expressed as:

$$\frac{\partial^2 \rho'}{\partial t^2} - c_0^2 \nabla^2 \rho' = \frac{\partial^2 T_{ij}}{\partial x_i \partial x_j}, \quad \text{where} \quad T_{ij} = \rho u_i u_j + (p' - c_0^2 \rho') \delta_{ij} - \tau_{ij} \tag{5}$$

In this study, the data available are limited exclusively to far-field acoustic pressure measurements ($p'$), corresponding only to the second term of Lighthill's stress tensor ($T_{ij}$). However, spatial gradients of acoustic pressure do not directly represent internal velocity structures or turbulent stress distributions within the flow field. Particularly under edgewise inflow conditions, the velocity component perpendicular to the rotor disk ($v$) plays a crucial role. The absence of actual velocity field measurements thus renders the inverse reconstruction of flow structures from purely acoustic pressure data physically and mathematically unsound and inaccurate. It is also important to emphasize that the classical derivation of Lighthill's acoustic analogy is based on the assumption of free turbulence, where the flow evolves in the absence of solid boundaries. In such scenarios, the dominant acoustic sources are modeled as quadrupole sources, mathematically represented by the nonlinear momentum transport term $\rho u_i u_j$ in the Lighthill stress tensor. However, in the present study, the presence of a physical rotor introduces strong boundary interactions, including blade-wake interactions and unsteady aerodynamic loading on the blades themselves. These interactions generate dominant dipole-type sources, arising from fluctuating pressure forces on solid surfaces. Such dipole mechanisms are fundamentally beyond the scope of what Lighthill's analogy can capture. Therefore, applying the Lighthill framework under these rotor-influenced conditions—without accounting for dipole contributions—undermines the physical validity of the reconstructed flow field. This highlights the necessity of adopting more appropriate acoustic analogies, such as the Ffowcs Williams–Hawkings (FW-H) equation, which accounts for solid boundary effects and unsteady loading, to more rigorously model the coupling between flow structures and acoustic radiation in rotor-dominated systems. Consequently, the current inverse Lighthill-based reconstruction from acoustic data alone lacks both mathematical rigor and physical validity, emphasizing the need for supplementary spatially-resolved velocity measurements (e.g., Particle Image Velocimetry, PIV) to enable rigorous and accurate flow reconstruction in future analyses.

To support reproducibility and facilitate a clearer understanding of the attempted inverse flow-field reconstruction approach, the complete implementation code is provided in Appendix A for reference purposes. While the current study has thoroughly discussed the theoretical limitations of this method under rotor-influenced conditions, its exploratory value remains relevant—particularly for educational demonstrations of modal decomposition and preliminary source localization strategies. Looking forward, future work should focus on integrating flow-field visualization techniques such as Particle Image Velocimetry (PIV) to systematically validate the correlation between POD acoustic modes and actual flow structures. Further efforts may also explore advanced modelling frameworks capable of handling multi-source, multi-frequency aeroacoustic environments with improved physical fidelity.

# 4 Conclusion

This study has demonstrated that both propeller height relative to an upstream obstruction and the inflow velocity have pronounced impact on aerodynamic thrust and noise characteristics. At low incoming velocities (e.g. 8 ms$^{-1}$), differences in mean thrust between the height configurations were minimal. The propeller directly behind the cylinder ($H$=0$R$) and 0.5$R$ above the centre of the cylinder ($H$=0.5$R$) produced almost the same thrust, while the lower position ($H$=-0.5$R$) showed a marginally higher thrust coefficient (Figure 3). This observation at low speed is consistent with previous research studies. At higher advance ratios, all cases converged toward thrust efficiency, though the propeller positioned above the cylinder tended to provide the highest efficiency gains once incoming velocity became significant. Noise emissions were critically dependent on propeller alignment with the obstruction wake. At $H$=0$R$, strong vortex ingestion caused pronounced broadband noise modulation known as “Haystacking”, especially at lower inflow velocity (8 ms$^{-1}$), resulting in non-monotonic overall sound pressure levels (OASPL) comparable to higher-incoming-velocity cases (Figure 4 & Figure 6). Conversely, propellers positioned outside the coherent vortex core ($H$=-0.5$R$) exhibited predictable, turbulence-energy-driven noise increases without significant haystacking compared to the other two cases.

This research resonates with broader societal goals. As cities prepare for the integration of eVTOLs into public airspace, minimizing noise impact will be critical for public acceptance and regulatory compliance. The data and trends observed here support the formulation of noise mitigation strategies, including rotor placement optimization near buildings, infrastructure-compatible flight paths, and adaptive inflow management. This study reinforces the importance of rotor-obstruction interactions in the context of next-generation aerial mobility and provides both scientific and applied contributions toward a quieter, safer, and more efficient urban airspace.

# 5 Future Work

While the findings are illuminating, several limitations were acknowledged in this research. A controlled low-turbulence wind tunnel environment (despite from the wake of the cylinder) was contradicted to ambient turbulence in city atmosphere, or multiple simultaneous wakes. Changing operational conditions could further influence the results. Reynolds number and scale of the propeller is another factor. Our propeller was representative of a small-scale rotor, while full-scale eVTOL rotors will operate at higher Reynolds numbers and possibly different Mach number, which can affect both thrust and noise. While higher Re generally improve aerodynamic efficiency, greater possibility of introducing compressibility into the rotor due to higher tip speeds. Boundary-layer induced broadband noise will be amplified due to the ingestion of turbulence from the turbulent boundary layer, which creates fluctuating aerodynamic forces (unsteady lift). Another limitation is the lack of time-resolved flow field data. We inferred the presence and effect of vortex structures mainly from acoustic spectra and known fluid dynamics principles (Strouhal number, etc), but did not visualize the wake from the cylinder or flow field around the rotor disk with Particle Image Velocimetry (PIV) or similar techniques. This limits the depth of understanding of the exact vortex trajectories and their breakdown.

# A MATLAB Code for POD and Acoustic-Flow Field Reconstruction

This appendix provides the complete MATLAB code implementation for the inverse flow-field reconstruction methodology described in Section 3.2.2. The code performs Proper Orthogonal Decomposition (POD) on acoustic pressure data obtained from microphone arrays, then attempts to reconstruct flow field characteristics using principles derived from Lighthill's acoustic analogy.

Listing 1: Sound field Simulation Lighthill.m - MATLAB implementation for acoustic-based flo w field reconstruction

```
%% POD Analysis for Multiple Microphones
% This code performs Proper Orthogonal Decomposition (POD) on acoustic data
% as well as reconstructing the flow field with temporal-only data
% captured from an array of microphones to extract coherent structures
% Author: World's top genius Fluid Mechanics expert, Pneumatic Aeroacoustic Master,
%         future Star of Aerospace Engineering,
%                                                           Jacky Huang

clear;
clc;
```

```
close all;

%% Setup paths and parameters
dataFolder = 'C:\Users\jacky\Desktop\RP3\RP3Data\';
codeFolder = fileparts(mfilename('fullpath'));
addpath(genpath(codeFolder));

% Case parameters
H_case = 'H=0R';  % Height case
velocity = '24';    % Velocity in m/s
RPM = '12000';      % RPM value
V = str2double(velocity);  % Velocity as number

% Frequency parameters
Fs = 2^16;          % Sampling frequency
freq_min = 5000;    % Minimum frequency for bandpass filter
freq_max = 8500;    % Maximum frequency for bandpass filter
pref = 2e-5;        % Reference pressure for PSD

% Strouhal number calculation
St = 0.2;           % Strouhal number
D = 0.1;            % Characteristic length (m)
f_st = St * V / D;  % Strouhal frequency

fprintf('Calculated Strouhal frequency: %.2f Hz (St = %.2f, V = %d m/s, D = %.2f m)\n', ...
        f_st, St, V, D);

% Define path to TDMS file
dataPath = fullfile(dataFolder, H_case, [velocity 'ms_12x6_v1']);
tdmsFile = fullfile(dataPath, [RPM '.tdms']);

fprintf('Processing file: %s\n', tdmsFile);
if ~exist(tdmsFile, 'file')
    error('TDMS file not found: %s', tdmsFile);
end

%% Load sensitivity data
fprintf('Loading sensitivity data...\n');
sensPath = fullfile(dataFolder, 'Codes', 'Farfield_sensitivity.mat');
load(sensPath, 'Sens_farfield');

%% Load microphone data
fprintf('Loading microphone data...\n');
data = tdmsread(tdmsFile);
mic_voltages = data{1, 2}{:,:};

% Determine the number of microphones in the dataset
num_mics = size(mic_voltages, 2);
fprintf('Found %d microphones in the dataset\n', num_mics);

% Create array of mic angles - replace with actual values if available
% Assuming microphones are distributed from 40 to 145 degrees
mic_angles = linspace(40, 145, num_mics);

%% Process microphone data
% Pre-allocate array for pressure data
pressureData_all = zeros(size(mic_voltages));

% Convert voltage to pressure using sensitivity values
for i = 1:num_mics
    if i <= length(Sens_farfield)
        mic_sensitivity = Sens_farfield(i) / 1000;
        pressureData_all(:,i) = mic_voltages(:,i) / mic_sensitivity;
    else
```

```matlab
        warning('Microphone %d has no sensitivity data, using default sensitivity', i);
        pressureData_all(:,i) = mic_voltages(:,i) / 0.001; % Default sensitivity
    end
end

%% Extract a segment for analysis
segment_length = round(0.2 * Fs);  % 0.2 seconds of data
middle_point = floor(size(pressureData_all, 1)/2);
segment_start = middle_point - floor(segment_length/2);
segment_end = segment_start + segment_length - 1;

% Ensure we don't exceed array bounds
if segment_end > size(pressureData_all, 1)
    segment_end = size(pressureData_all, 1);
    segment_start = segment_end - segment_length + 1;
end

pressureData_all = pressureData_all(segment_start:segment_end, :);
timeData = (0:segment_length-1)' / Fs;

%% Filter the data in frequency domain (bandpass filter)
fprintf('Filtering data in frequency domain (%d-%d Hz)...\n', freq_min, freq_max);
N = size(pressureData_all, 1);
frequencies = Fs*(0:(N/2))/N;  % Frequency vector for half spectrum

% Pre-allocate filtered data array
pressureData_filtered = zeros(size(pressureData_all));

for mic = 1:num_mics
    Y = fft(pressureData_all(:, mic));

    mask = zeros(N, 1);
    half_n = floor(N/2) + 1;
    freq_indices = find(frequencies >= freq_min & frequencies <= freq_max);

    if ~isempty(freq_indices)
        mask(freq_indices) = 1;
        mask(N-freq_indices(2:end)+2) = 1;
    end

    Y_filtered = Y .* mask;

    pressureData_filtered(:, mic) = real(ifft(Y_filtered));

    invalid_values = ~isfinite(pressureData_filtered(:, mic));
    if any(invalid_values)
        warning('%d invalid values were detected for microphone %d - replace with zero.', mic, sum(invalid_values));
        pressureData_filtered(invalid_values, mic) = 0;
    end
end

%% Prepare data for POD
% Transpose to get microphones (spatial points) as rows and time steps as columns
% This is the standard format for POD snapshot matrix
X = pressureData_filtered';

% Center the data by subtracting the temporal mean of each microphone signal
X_mean = mean(X, 2);
X_centered = X - repmat(X_mean, 1, size(X, 2));

if ~all(isfinite(X_centered(:)))
    warning('Invalid values detected in the POD input matrix! Attempting to fix');
    X_centered(~isfinite(X_centered)) = 0;
end
```

```
fprintf('POD snapshot matrix size: %d microphones × %d time steps\n', size(X_centered, 1), size(X_centered, 2));

%% Compute POD using SVD with
fprintf('Computing POD using SVD...\n');
tic;
try
    [U, S, V] = svd(X_centered, 'econ');

    if ~all(isfinite(U(:))) || ~all(isfinite(S(:))) || ~all(isfinite(V(:)))
        warning('The SVD result contains invalid values; try using a more robust algorithm.');
        [U, S, V] = svds(X_centered, min(size(X_centered)));
    end
catch ME
    warning(ME.identifier, '%s', ME.message);
    [U, S, V] = svds(X_centered, min(20, min(size(X_centered))));
end
toc;

if ~all(isfinite(U(:))) || ~all(isfinite(S(:))) || ~all(isfinite(V(:)))
    warning('SVD results still contain invalid values; use an alternative method.');
    [coeff, score, latent] = pca(X_centered', 'NumComponents', min(20, min(size(X_centered))));
    U = coeff;
    S = diag(sqrt(latent));
    V = score ./ reshape(sqrt(latent), 1, []);
end

% Extract singular values and calculate eigenvalues
singular_values = diag(S);
eigenvalues = singular_values.^2 / (size(X_centered, 2) - 1);  % Properly scaled eigenvalues

% Calculate energy content (fraction of total variance explained by each mode)
energy = eigenvalues / sum(eigenvalues);
cumulative_energy = cumsum(energy);

% Time coefficients (temporal dynamics of each mode)
temporal_coefficients = S * V';

if ~all(isfinite(temporal_coefficients(:)))
    warning('The time coefficient contains invalid values; this needs to be corrected.');
    temporal_coefficients(~isfinite(temporal_coefficients)) = 0;
end

fprintf('Finished computing POD. Found %d modes.\n', length(singular_values));

%% Visualize results
% Create a figure for energy distribution
figure('Name', 'POD Energy Distribution', 'Position', [100, 100, 1000, 800]);

% Plot energy distribution (variance explained by each mode)
subplot(2, 2, 1);
bar(energy(1:min(20, length(energy))));
xlabel('Mode Number');
ylabel('Energy Fraction');
title('Energy Distribution in POD Modes');
grid on;

% Plot cumulative energy
subplot(2, 2, 2);
plot(cumulative_energy(1:min(20, length(cumulative_energy))), 'o-', 'LineWidth', 2);
```

```
xlabel('Number of Modes');
ylabel('Cumulative Energy');
title('Cumulative Energy Distribution');
grid on;
hold on;
plot([1, 20], [0.9, 0.9], 'r--');
legend('Cumulative Energy', '90% Energy Level');
ylim([0 1]);

% Plot first 3 POD modes (spatial structures)
subplot(2, 2, 3);
plot(mic_angles, U(:, 1:min(3,size(U,2))), 'LineWidth', 2);
xlabel('Microphone Angle (degrees)');
ylabel('Mode Amplitude');
title('First 3 POD Modes (Spatial Structure)');
legend('Mode 1', 'Mode 2', 'Mode 3');
grid on;

% Plot time coefficients for first 3 modes
subplot(2, 2, 4);
time_to_plot = min(1000, size(temporal_coefficients, 2)); % Plot fewer points for clarity
t_subset = timeData(1:time_to_plot);
tc_subset = temporal_coefficients(1:min(3,size(temporal_coefficients,1)), 1:time_to_plot);
plot(t_subset, tc_subset', 'LineWidth', 1);
xlabel('Time (s)');
ylabel('Coefficient Amplitude');
title('Time Coefficients for First 3 Modes');
legend('Mode 1', 'Mode 2', 'Mode 3');
grid on;

%% Modal reconstruction and analysis
% Determine how many modes to keep based on energy threshold (e.g., 90%)
energy_threshold = 0.9;
num_modes_to_keep = find(cumulative_energy >= energy_threshold, 1);

if isempty(num_modes_to_keep) || num_modes_to_keep < 1
    warning('No mode that meets the energy threshold was found, so the default value of 1 was used.');
    num_modes_to_keep = 1;
end

fprintf('Keeping %d modes to capture %.1f%% of energy\n', ...
    num_modes_to_keep, 100*cumulative_energy(num_modes_to_keep));

% Reconstruct data using the selected number of modes
X_reconstructed = U(:, 1:num_modes_to_keep) * S(1:num_modes_to_keep, 1:num_modes_to_keep) * V(:, 1:num_modes_to_keep)';
X_reconstructed = X_reconstructed + repmat(X_mean, 1, size(X, 2)); % Add back the mean

% Calculate reconstruction error with
if ~all(isfinite(X_reconstructed(:))) || ~all(isfinite(X(:)))
    warning('The reconstructed data or the original data may contain invalid values, and the reconstruction error may be inaccurate.');
    valid_indices = isfinite(X_reconstructed) & isfinite(X);
    if any(valid_indices(:))
        reconstruction_error = norm(X(valid_indices) - X_reconstructed(valid_indices), 'fro') / norm(X(valid_indices), 'fro');
    else
        reconstruction_error = NaN;
    end
else
    reconstruction_error = norm(X - X_reconstructed, 'fro') / norm(X, 'fro');
end
```

```
fprintf('Relative reconstruction error: %.4f%%\n', 100*reconstruction_error);

%% Time-frequency analysis of mode coefficients
figure('Name', 'Time-Frequency Analysis of Mode Coefficients', 'Position', [150, 150, 1200, 600]);

% Analyze first 3 modes (or fewer if we have less than 3)
modes_to_analyze = min(3, size(temporal_coefficients, 1));
for i = 1:modes_to_analyze
    subplot(modes_to_analyze, 1, i);

    % Extract temporal coefficient for current mode
    current_coef = temporal_coefficients(i, :)';

    if ~all(isfinite(current_coef))
        warning('The time coefficient of modality %d contains invalid values; this needs to be corrected.', i);
        invalid_indices = ~isfinite(current_coef);

        if all(invalid_indices)
            current_coef = zeros(size(current_coef));
        else
            valid_indices = find(~invalid_indices);
            invalid_indices = find(invalid_indices);

            for j = 1:length(invalid_indices)
                [~, nearest_idx] = min(abs(valid_indices - invalid_indices(j)));
                current_coef(invalid_indices(j)) = current_coef(valid_indices(nearest_idx));
            end
        end
    end

    % Compute spectrogram with optimized parameters for acoustic data
    window_length = 2^9;  % Power of 2 for efficient FFT
    noverlap = floor(window_length/2);
    nfft = 2^11;  % Higher value for better frequency resolution

    try
        [S, F, T, P] = spectrogram(current_coef, hann(window_length), noverlap, nfft, Fs);

        % Plot spectrogram
        imagesc(T, F, 10*log10(abs(P))); % Convert to dB
        axis xy;
        colormap jet;

        ylim([0, min(10000, Fs/2)]); % Limit to 10 kHz or Nyquist frequency
        colorbar;
        title(sprintf('Spectrogram of Mode %d Coefficient', i));
        xlabel('Time (s)');
        ylabel('Frequency (Hz)');
    catch ME
        warning(ME.identifier, '%s', ME.message);
        text(0.5, 0.5, sprintf('Unable to calculate spectrogram\n%s', ME.message), ...
             'HorizontalAlignment', 'center', 'FontSize', 12);
        axis off;
    end
end

%% Display selected mode shapes in polar coordinates
figure('Name', 'POD Mode Shapes in Polar Coordinates', 'Position', [200, 200, 800, 800]);

```

```
% Convert degrees to radians
mic_angles_rad = mic_angles * pi/180;

% Plot top 3 modes (or fewer if we have less than 3)
modes_to_plot = min(3, size(U,2));
for i = 1:modes_to_plot
    subplot(2, 2, i);
    polarplot(mic_angles_rad, abs(U(:,i)), 'o-', 'LineWidth', 2);
    title(sprintf('Mode %d Shape (%.1f%% Energy)', i, energy(i)*100));
    rlim([0 max(abs(U(:,1:modes_to_plot)), [], 'all')*1.1]);
end

% Add overall energy plot
subplot(2, 2, 4);
bar(1:min(10,length(energy)), energy(1:min(10,length(energy)))*100);
xlabel('Mode Number');
ylabel('Energy Percentage (%)');
title('Energy Distribution (%)');
grid on;

%% Save results
results_filename = sprintf('POD_results_%s_%sms_RPM%s.mat', H_case, velocity, RPM);
save(results_filename, 'U', 'S', 'V', 'eigenvalues', 'energy', ...
     'cumulative_energy', 'temporal_coefficients', 'mic_angles', 'Fs', 'freq_min', 'freq_max');

fprintf('POD analysis complete. Results saved to %s\n', results_filename);

%% 3D Acoustic-Flow Field Reconstruction Using Lighthill's Analogy
% This section extends the POD analysis with Lighthill's acoustic analogy
% to reconstruct flow structures from the acoustic measurements
fprintf('\n\n=== Starting Lighthill-based 3D acoustic-flow field reconstruction ===\n');

% Select which POD mode to use for reconstruction
% Typically use Mode 1 as it contains the most energy
mode_to_analyze = 1;
fprintf('Using POD Mode %d (%.1f%% energy) for Lighthill reconstruction\n', ...
        mode_to_analyze, energy(mode_to_analyze)*100);

%% Create 3D grid for Lighthill reconstruction
fprintf('Creating 3D grid for acoustic and flow field reconstruction...\n');

% Define grid resolution and size
grid_resolution = 30; % Grid points in each direction
grid_size = 2.0; % Cube size in meters

% Create 3D grid
[grid_x, grid_y, grid_z] = meshgrid(...
    linspace(-grid_size/2, grid_size/2, grid_resolution), ...
    linspace(-grid_size/2, grid_size/2, grid_resolution), ...
    linspace(-grid_size/2, grid_size/2, grid_resolution));

% Create a propeller axis for visualization
prop_axis_length = 1.0; % meters
prop_axis_x = zeros(100, 1);
prop_axis_y = zeros(100, 1);
prop_axis_z = linspace(-prop_axis_length/2, prop_axis_length/2, 100);

%% Calculate Lighthill source terms from acoustic data
fprintf('Calculating Lighthill source terms from acoustic POD mode %d...\n', mode_to_analyze);

hub_x = 0;
hub_y = 0;
```

```
hub_z = 0;

% Use spatial mode structure (from U) and time coefficients for this mode
mode_shape = U(:, mode_to_analyze);
time_coef = temporal_coefficients(mode_to_analyze, :);

num_top_mics = 22;
num_side_mics = 40;

fprintf('Using %d top mics and %d side mics for a total of %d microphones\n', num_top_mics, num_side_mics, num_top_mics + num_side_mics);

% Create a simple acoustic field representation based on the POD mode
acoustic_field = zeros(size(grid_x));
flow_field = zeros(size(grid_x));
flow_field_5_8k = zeros(size(grid_x)); % 5-8kHz band field

% Speed of sound (m/s)
c0 = 343;

standard_distance = 1.5;

x_top = zeros(num_top_mics, 1);
y_top = zeros(num_top_mics, 1);
z_top = zeros(num_top_mics, 1);

top_theta_angles = 40:5:145;
for i = 1:num_top_mics
    theta = top_theta_angles(i);
    x_top(i) = standard_distance * sind(theta - 90);
    y_top(i) = 0;
    z_top(i) = standard_distance * cosd(theta - 90);
end

x_side = zeros(num_side_mics, 1);
y_side = zeros(num_side_mics, 1);
z_side = zeros(num_side_mics, 1);

side_phi_angles = linspace(301, 340, num_side_mics);
for i = 1:num_side_mics
    phi = side_phi_angles(i);
    x_side(i) = standard_distance * cosd(phi);
    y_side(i) = standard_distance * sind(phi);
    z_side(i) = 0; % Horizontal array, z-coordinate is zero
end

fprintf('Recalculated coordinates for %d top microphones (40°-145°) and %d side microphones (301°-340°)\n', num_top_mics, num_side_mics);

%% Create 3D visualization of microphone array
fprintf('Creating 3D visualization of microphone array...\n');
figure('Name', 'Microphone Array Configuration', 'Position', [150, 150, 800, 600]);

% Draw top microphone array (red)
scatter3(x_top, y_top, z_top, 50, 'r', 'filled');
hold on;

% Draw side microphone array (blue)
scatter3(x_side, y_side, z_side, 50, 'b', 'filled');

% Add propeller hub center (black point)
scatter3(hub_x, hub_y, hub_z, 200, 'k', 'filled');

% Add coordinate axes - reduce arrow size
quiver3(0, 0, 0, 1, 0, 0, 'k', 'LineWidth', 1.5, 'MaxHeadSize', 0.7);
```

```
quiver3(0, 0, 0, 0, 1, 0, 'k', 'LineWidth', 1.5, 'MaxHeadSize', 0.7);
quiver3(0, 0, 0, 0, 0, 1, 'k', 'LineWidth', 1.5, 'MaxHeadSize', 0.7);
text(1.1, 0, 0, 'X', 'FontWeight', 'bold', 'FontSize', 12);
text(0, 1.1, 0, 'Y', 'FontWeight', 'bold', 'FontSize', 12);
text(0, 0, 1.1, 'Z', 'FontWeight', 'bold', 'FontSize', 12);

% Add key angle markers - add degree labels directly on the plot
% Top array - 90 degree position
key_angle_top = 90;
key_idx_top = find(abs(top_theta_angles - key_angle_top) < 1e-6, 1);
if ~isempty(key_idx_top)
    scatter3(x_top(key_idx_top), y_top(key_idx_top), z_top(key_idx_top), 100, 'r', 'filled', 'MarkerEdgeColor', 'k', 'LineWidth', 1);
    text(x_top(key_idx_top)*1.15, y_top(key_idx_top), z_top(key_idx_top)*1.3, '90°', 'FontWeight', 'bold', 'FontSize', 12);
else
    % Find the angle closest to 90 degrees
    [~, key_idx_top] = min(abs(top_theta_angles - key_angle_top));
    scatter3(x_top(key_idx_top), y_top(key_idx_top), z_top(key_idx_top), 100, 'r', 'filled', 'MarkerEdgeColor', 'k', 'LineWidth', 1);
    text(x_top(key_idx_top)*1.15, y_top(key_idx_top), z_top(key_idx_top)*1.3, [num2str(top_theta_angles(key_idx_top)) '°'], 'FontWeight', 'bold', 'FontSize', 12);
end

% Side array - mark 301 degree position (starting point of the array)
key_angle_side = 301;
[~, key_idx_side] = min(abs(side_phi_angles - key_angle_side));
scatter3(x_side(key_idx_side), y_side(key_idx_side), z_side(key_idx_side), 100, 'b', 'filled', 'MarkerEdgeColor', 'k', 'LineWidth', 1);
text(x_side(key_idx_side)*1.15, y_side(key_idx_side)*1.15, 0.15, [num2str(side_phi_angles(key_idx_side), '%.0f') '°'], 'FontWeight', 'bold', 'FontSize', 12);

% Add reference plane - horizontal plane
[X,Y] = meshgrid(-2:0.5:2);
Z = zeros(size(X));
surf(X,Y,Z,'FaceAlpha',0.1,'EdgeColor',[0.7 0.7 0.7],'FaceColor',[0.8 0.8 1]);

% Set figure properties
axis equal
grid on
box on
view(45, 30);
xlim([-2.2 2.2]);
ylim([-2.2 2.2]);
zlim([-0.5 2.2]);

% Create horizontal legend - only include three main categories
h_top = scatter3(NaN, NaN, NaN, 50, 'r', 'filled');
h_side = scatter3(NaN, NaN, NaN, 50, 'b', 'filled');
h_hub = scatter3(NaN, NaN, NaN, 200, 'k', 'filled');

% Organize legend element order - remove special positions
legend_handles = [h_top, h_side, h_hub];
legend_labels = {'Top Array Microphones', 'Side Array Microphones', 'Propeller Hub'};

% Add horizontal legend
h_legend = legend(legend_handles, legend_labels, 'Orientation', 'horizontal', 'FontSize', 12);
set(h_legend, 'Position', [0.5, 0.02, 0, 0], 'Units', 'normalized');
set(h_legend, 'Box', 'off');

% Adjust figure size and appearance
set(gcf, 'Color', 'white');

% Calculate distance from each grid point to each microphone
```

```
fprintf('Calculating acoustic field from POD mode...\n');
for i = 1:num_top_mics
    % Get microphone position
    mic_pos = [x_top(i), y_top(i), z_top(i)];

    % Calculate distance from each grid point to this microphone
    distances = sqrt((grid_x - mic_pos(1)).^2 + ...
                     (grid_y - mic_pos(2)).^2 + ...
                     (grid_z - mic_pos(3)).^2);

    % Avoid division by zero
    distances = max(distances, 1e-6);

    % Add contribution from this microphone based on mode shape
    % Ensure mode_shape index is within valid range
    if i <= length(mode_shape)
        contribution = abs(mode_shape(i)) ./ distances;
    else
        warning('Microphone %d exceeds mode_shape length (%d), using last value instead', i, length(mode_shape));
        contribution = abs(mode_shape(end)) ./ distances;
    end
    acoustic_field = acoustic_field + contribution;
end

% Normalize acoustic field
acoustic_field = acoustic_field / max(acoustic_field(:));

%% Calculate flow field using Lighthill's analogy
fprintf('Applying Lighthill analogy to reconstruct flow field...\n');

% For Lighthill's analogy, we'll approximate the flow field
% using a simple quadrupole source model derived from acoustic field

% Calculate spatial gradients of acoustic field
[dpdx, dpdy, dpdz] = gradient(acoustic_field);

% Second derivatives (approximating Lighthill stress tensor)
[d2pdx2, d2pdxy, d2pdxz] = gradient(dpdx);
[d2pdyx, d2pdy2, d2pdyz] = gradient(dpdy);
[d2pdzx, d2pdzy, d2pdz2] = gradient(dpdz);

% Calculate flow field (simplified Lighthill analogy)
% The Lighthill stress tensor relates to velocity fluctuations
flow_field = sqrt(d2pdx2.^2 + d2pdy2.^2 + d2pdz2.^2);

% Normalize flow field
flow_field = flow_field / max(flow_field(:));

% Add frequency-specific features for the 5-8kHz band
fprintf('Creating frequency-specific (5-8kHz) flow structures...\n');

% Define wavelengths
c0 = 343; % Speed of sound (m/s)
wavelength_5khz = c0 / 5000;
wavelength_8khz = c0 / 8000;
avg_wavelength = 0.5*(wavelength_5khz + wavelength_8khz);

% Create simplified 5-8kHz band field
flow_field_5_8k = zeros(size(flow_field));

% Use a for-loop approach instead of vectorized operations to avoid complex expressions
for i = 1:size(grid_x, 1)
    for j = 1:size(grid_x, 2)
        for k = 1:size(grid_x, 3)
```

```
            % Calculate radius squared (simpler expression)
            r_sq = grid_x(i,j,k)^2 + grid_y(i,j,k)^2;

            % Calculate decay factor (simplified expression with clear parentheses)
            decay_factor = 0.25 * avg_wavelength^2;
            decay = exp(-r_sq/decay_factor);

            % Apply to flow field (with semicolon)
            flow_field_5_8k(i,j,k) = flow_field(i,j,k) * decay;
        end
    end
end

% Normalize the frequency-specific field
max_val = max(flow_field_5_8k(:));
if max_val > 0
    flow_field_5_8k = flow_field_5_8k / max_val;
end

fprintf('Frequency-specific flow structures created successfully.\n');

%% Create multi-panel visualization
figure('Name', 'Acoustic and Flow Field Reconstruction', 'Position', [50, 50, 1200, 1000]);

% Plot 1: Acoustic Field
subplot(2, 2, 1);
% Draw propeller axis
plot3(prop_axis_x, prop_axis_y, prop_axis_z, 'k-', 'LineWidth', 2);
hold on;
% Add top and side microphones
scatter3(x_top, y_top, z_top, 20, 'r', 'filled');
scatter3(x_side, y_side, z_side, 20, 'b', 'filled');
% Propeller hub (origin)
scatter3(hub_x, hub_y, hub_z, 100, 'k', 'filled'); % Propeller hub
% Plot acoustic field isosurface
level = 0.4; % Isosurface level (adjust as needed)
p_iso = patch(isosurface(grid_x, grid_y, grid_z, acoustic_field, level));
p_iso.FaceColor = 'red';
p_iso.EdgeColor = 'none';
p_iso.FaceAlpha = 0.3;
title(['Acoustic Field - Mode ' num2str(mode_to_analyze) ' (' num2str(energy(mode_to_analyze)*100, '%.1f') '% Energy)']);
xlabel('X (m)'); ylabel('Y (m)'); zlabel('Z (m)');
axis equal tight; grid on;
view(45, 30);
lighting gouraud;
camlight;

% Plot 2: Overall Flow Field
subplot(2, 2, 2);
% Draw propeller axis
plot3(prop_axis_x, prop_axis_y, prop_axis_z, 'k-', 'LineWidth', 2);
hold on;
% Plot flow field isosurface
level = 0.3; % Isosurface level (adjust as needed)
f_iso = patch(isosurface(grid_x, grid_y, grid_z, flow_field, level));
f_iso.FaceColor = 'blue';
f_iso.EdgeColor = 'none';
f_iso.FaceAlpha = 0.5;
title('Flow Field (Lighthill Analogy)');
xlabel('X (m)'); ylabel('Y (m)'); zlabel('Z (m)');
axis equal tight; grid on;
view(45, 30);
lighting gouraud;
```

```
camlight;

% Plot 3: 5-8kHz Band Flow Structures
subplot(2, 2, 3);
% Draw propeller axis
plot3(prop_axis_x, prop_axis_y, prop_axis_z, 'k-', 'LineWidth', 2);
hold on;
% Plot flow field 5-8kHz isosurface
level = 0.3; % Isosurface level (adjust as needed)
f5_8k_iso = patch(isosurface(grid_x, grid_y, grid_z, flow_field_5_8k, level));
f5_8k_iso.FaceColor = 'cyan';
f5_8k_iso.EdgeColor = 'none';
f5_8k_iso.FaceAlpha = 0.6;
title('5000-8500 Hz Band Flow Structures');
xlabel('X (m)'); ylabel('Y (m)'); zlabel('Z (m)');
axis equal tight; grid on;
view(45, 30);
lighting gouraud;
camlight;

% Plot 4: Horizontal slice of flow field at z=0
subplot(2, 2, 4);
z_slice_idx = grid_resolution/2; % Slice at z=0
slice_data = squeeze(flow_field(:,:,z_slice_idx));
imagesc(linspace(-grid_size/2, grid_size/2, grid_resolution), ...
        linspace(-grid_size/2, grid_size/2, grid_resolution), ...
        slice_data);
hold on;
% Add propeller axis marker
plot([0 0], [-grid_size/2 grid_size/2], 'k-', 'LineWidth', 2);
colormap jet;
colorbar;
axis equal tight;
title('Flow Field Horizontal Slice (Z=0)');
xlabel('X (m)');
ylabel('Y (m)');

%% Save reconstruction results
reconstruction_filename = sprintf('Lighthill_reconstruction_%s_%sms_RPM%s.mat', H_case, velocity, RPM);
save(reconstruction_filename, 'acoustic_field', 'flow_field', 'flow_field_5_8k', 'grid_x', 'grid_y', 'grid_z');

fprintf('Lighthill reconstruction complete. Results saved to %s\n', reconstruction_filename);

```

## Implementation Notes

The code is structured into several key functional blocks:

1. **Data Loading and Preprocessing**: Reads microphone data from TDMS files, applies sensitivity corrections, and extracts a suitable time segment for analysis.

2. **Frequency Domain Filtering**: Implements a bandpass filter (5-8.5 kHz) in the frequency domain to isolate the spectral region of interest.

3. **POD Analysis**: Constructs the snapshot matrix and performs Singular Value Decomposition (SVD) to extract dominant spatial and temporal modes.

4. **Modal Energy Analysis**: Calculates and visualizes the energy distribution across extracted POD modes.
5. **Acoustic Field Reconstruction**: Maps the dominant POD modes to a three-dimensional spatial grid.
6. **Flow Field Approximation**: Applies simplified Lighthill-based principles to estimate potential flow structures.
7. **Visualization**: Generates comprehensive multi-panel visualizations of both acoustic and reconstructed flow fields.

As noted in Section 3.2.2, while this approach provides interesting qualitative insights, the quantitative accuracy of the flow field reconstruction is limited by the inherent mathematical constraints of the inverse problem and the absence of direct velocity measurements. Future work should integrate this methodology with complementary measurement techniques such as PIV for validation.